\documentclass[10pt]{wlscirep}
\usepackage[utf8]{inputenc}
\usepackage[T1]{fontenc}
\usepackage{amsmath}
\usepackage{float}
\title{A Renewable Double Plasma Mirror For Petawatt-class Lasers}

\author[1,*]{Nick Czapla}
\author[2]{Derek M. Nasir}
\author[3]{Lieselotte Obst-Huebl}
\author[2]{Anthony Zingale}
\author[3,a]{Jianhui Bin}
\author[3]{Anthony J. Gonsalves}
\author[3,b]{Sven Steinke}
\author[3]{Kei Nakamura}
\author[3]{Carl B. Schroeder}
\author[3]{Eric Esarey}
\author[3]{Cameron G. R. Geddes}
\author[2]{Douglass W. Schumacher}
\affil[1]{SLAC National Accelerator Laboratory, Menlo Park, CA 94025, USA}
\affil[2]{Physics Department, The Ohio State University, Columbus, OH 43210, USA}
\affil[3]{Accelerator Technology and Applied Physics Division, Lawrence Berkeley National Laboratory, Berkeley, CA 94720, USA}

\affil[*]{Correspondence and requests for materials should be addressed to N.C. (nczapla@slac.stanford.edu)}
\affil[a]{\textbf{Present address:} State Key Laboratory of High Field Laser Physics and CAS Center for Excellence in Ultra-Intense Laser Science, Shanghai Institute of Optics and Fine Mechanics, Chinese Academy of Sciences, Shanghai 201800, China.}
\affil[b] {\textbf{Present address:} Marvel Fusion Gmbh, blumenstrasse 28, 80331 Munich, Germany.}

\begin{abstract}
Exceptional pulse contrast can be critical for ultraintense laser experiments, particularly when using solid density targets, and their use is becoming widespread. However, current plasma mirror technology is becoming inadequate for the new generation of high repetition rate, high power lasers now available. We describe a novel double plasma mirror configuration based on renewable, free standing, ultrathin liquid crystal films tested at the BELLA Petawatt Laser Center. Although operating at a repetition rate of several shots per minute, this system can be scaled to a high repetition rate exceeding 1 Hz and represents an important step towards enabling sustained, continuous operation of plasma mirrors. We demonstrate an improvement of two to three orders of magnitude in contrast and a total throughput of ~80\%. We present the first measurements of a beam reflected from a single or double plasma mirror system using a wavefront sensor, showing a well preserved wavefront and spatial mode. Finally, we introduce a model that predicts the total throughput through this double plasma mirror. This is the first model that accurately predicts the peak reflectivity of a plasma mirror when given the laser temporal profile.
\end{abstract}
\begin{document}

\flushbottom
\maketitle
%
%
\thispagestyle{empty}

\section*{Introduction}

State-of-the-art high power laser facilities can readily achieve intensities $>10^{21}$ W/cm$^{2}$ and intensities $>10^{23}$ W/cm$^{2}$ are now coming online\cite{Yoon:21}. 
At these extreme intensities, inevitable pre-pulses and pedestals preceding the main pulse must be considered when designing an experiment\cite{Hadjisolomou:10.1063/1.5124457}. 
For example, at $10^{21}$ W/cm$^{2}$, $10^{-8}$ contrast pre-pulses (pre-pulses with intensity a factor of $10^{8}$ below the peak of the main pulse) would result in ionization and heating of the target before the main pulse arrives. 
On the other hand, high-contrast laser-ion acceleration has resulted in significantly increased ion energies when using thin targets\cite{Hadjisolomou:10.1063/1.5124457,Hegelich_2013,10.1063/1.2220011,10.1063/1.2480610,PhysRevLett.99.185002,RevModPhys.85.751}. 
In any case, pre-plasma due to pre-pulses and slow pedestal turn-on can substantially modify a target and complicate the interpretation of an experiment\cite{Zingale_thesis}\cite{Pozderac_thesis}.
\par
Single or dual plasma mirrors (PMs) in the laser beamline are commonly used to improve contrast by several orders of magnitude at the cost of a loss of pulse energy\cite{10.1063/1.4954242,Thaury_2007,Scott_2015,Rodel_2011,Obst_2018,osti_1349103,PhysRevE.69.026402,Inoue:16,10.1063/1.1646737,osti_21272732,Nomura_2007,10.1063/1.5109683,Monot:04}. 
Typically, a PM is an anti-reflection coated substrate that transmits light below the ionization threshold of the coating. 
After the intensity of the pulse crosses this threshold, the PM ionizes and becomes reflective. 
This self-triggering mechanism results in a very fast acting mirror that, ideally, only reflects the peak of the pulse and post-pulses, transmitting the pedestal and pre-pulses out of the beamline. 
Double plasma mirror (DPM) systems further enhance the contrast at the cost of still more pulse energy lost\cite{10.1063/1.2234850,Levy:07,Kim_2011}. 
As PM operation is inherently destructive to the substrate, PMs are often implemented using a large substrate that is rastered to a fresh location after each shot. 
This size is a limiting factor to the total number of shots possible before needing to replace the substrate, usually requiring a venting/pumping cycle of a vacuum chamber. 
Moreover, the higher the laser power, the larger the laser spot size required on the PM because of the need to carefully control the time during the pulse when the ionization threshold is crossed. 
For this reason, current PM technology is challenged by sustained operation of high-power, high repetition rate laser systems. 
Finally, it is also imperative that the PM or DPM enhances the temporal contrast without sacrificing spatial mode quality, pointing or too much of the energy available on target. 
Here we report on substantial progress towards PMs and DPMs that can satisfy these constraints based on free-standing, ultrathin liquid crystal films.
\par
Previously we reported on a novel PM constructed using the liquid crystal (LC) 8CB (4-octyl-4’-cyanobiphenyl) as a free standing film formed in situ before every shot\cite{osti_1349103}. 
These LC films can reliably form stable films <40 nm for which the reflectivity is low due to thin film interference and, thus, do not require an AR coating. 
Using a Linear Slide Target Inserter (LSTI)\cite{osti_1349103}, single 8CB films have been used to reflect up to 75\% of the incident light with a maximum possible contrast enhancement exceeding two orders of magnitude. 
Since the LSTI forms each PM film in the same location in situ before every shot, there is no need to raster optics or replace an underlying substrate. 
Accumulation of debris on the PM surface and nearby optics is lower when liquid crystal films are used due to the low volume of the films (nanoliters or less). 
Moreover, the PM surface is cleaned by the same wiper that forms the films. 
It was recently found that LC PMs also make quality beam diverters and can, for example, be used to separate a laser pulse from an accelerated electron beam with little degradation to the emittance\cite{PhysRevAccelBeams.24.121301}.
\par
Here we present the results of a novel DPM system fabricated using free standing LC films with on-shot characterization of the temporal contrast enhancement at both the nano- and picosecond timescales, total reflection, wavefront quality, spatial mode, and system pointing. 
This work was performed using the Berkeley Lab Laser Accelerator (BELLA) Center PW laser\cite{7934119}. 
At optimal performance, we observed a total throughput >80\% and a maximum contrast enhancement of 2-3 orders of magnitude. 
We also present a novel model that accurately predicts the total reflection over a large range of intensities ($10^{14}$ to $>10^{18}$ W/cm$^{2}$). 
This model is validated using the new results here and a previous LC PM experiment. 
It is the first to accurately predict both the peak reflectivity and the intensity at which reflectivity falls off, thus establishing the operating range of a PM.

\section*{Experimental Setup}

The petawatt (PW) laser at the BELLA center is a Ti:sapphire double chirped pulse amplification (CPA) based system that delivers a 35 fs pulse with a peak power of 1.2 PW, although we ran at lower power, 0.2 PW. 
It is the first such system to be able to operate at a 1 Hz repetition rate \cite{7934119} and it is part of the LaserNetUS network, under which this work was performed. 
For this work, a long focal length, F/65, off-axis parabolic (OAP) mirror focused the beam to an intensity FWHM spot size of 65 \textmu m. 
The experiment was performed in vacuum and the setup is shown in Fig. 1 with the two PMs labeled “PM 1” and “PM 2”. 
After the OAP, the beam was incident on PM 1 with p-polarization at an angle of incidence of ~8\textdegree. 
The reflected beam was incident onto PM 2 at an angle of incidence of ~11\textdegree. 
The beam could also bypass the DPM for comparison of beam characteristics with and without the DPM. 
As will be discussed in more detail later, there is an optimal intensity at which PMs perform, ranging between $10^{15}$ W/cm$^{2}$ and $10^{17}$ W/cm$^{2}$ depending on the PM substrate used and the pre-pulse and pedestal present. 
Our previous experiments indicated that for liquid crystal films, this is typically ~2 x $10^{16}$ W/cm$^{2}$\cite{osti_1349103}.  
For a given temporal contrast, power, and spatial profile there is an optimal spot size, larger for higher power systems. 
During DPM characterization a deformable mirror (DM) was utilized to move the focus along the optical axis allowing a scan of the location of the beam waist in the DPM through a range exceeding 150 mm with negligible change of the waist size. 
The characterization and optimization were performed using 7 J pulses incident on PM 1. A restricted set of measurements was also performed using a pulse energy of 20 J.

\begin{figure}[h]
\centering
\includegraphics[width=0.9\linewidth]{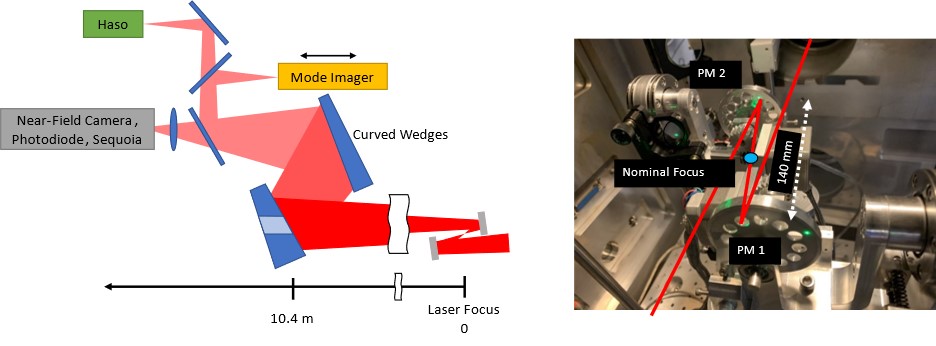}
\caption{Experimental layout of DPM system shown on the right. PM 1 and PM 2 are on gimbal mounts for individual angle alignment. The beam waist was positioned along the optical axis by using the DM. The focal scan ranged from >60 mm upstream of PM 2 to >60 mm downstream of PM 2. The angle alignment of each PM was verified using a CW laser diode alignment beam. Corrections to the beam pointing were performed using motorized gimbal mounts. A schematic of the diagnostic used is shown on the left. The sizes are not to scale.}
\label{fig:Setup}
\end{figure}
\par
Each PM system was based on the Spinning Disk Inserter (SDI) \cite{Zingale_thesis}\cite{Schumacher_2017}, a rotary version of the LSTI, producing film thicknesses ranging from 20 to 40 nm. 
Briefly, an SDI forms films using a spinning wheel with 12 x 8 mm apertures that sweep past a fixed wiper made of an absorbent material saturated with 8CB. 
The film thickness is dependent on the rotation speed and temperature. 
Each SDI was mounted onto a specially designed gimbal mount that allowed for pitch and yaw control around the center of the PM aperture. 
Before every shot, the alignment of each film was verified independently using the reflection of a 532 nm CW laser diode beam that was split into two beams, which were reflected of both films in separate alignment beamlines (Fig. 1). 
A computer operated pointing feedback system controlled a set of vacuum motors (Picomotors, Newfocus) and adjusted the pitch and yaw of each gimbal mount until the alignment laser reflections were in a selected position on a screen. 
This procedure resulted in a pointing accuracy through the entire DPM of ~0.4 mrad, better than the 0.5 mrad angle acceptance of the high power laser diagnostics employed, which were located ~13 m downstream from the DPM. 
The total combined time for wiping and aligning new PM films typically ranged from 20 to 30 seconds. 
Film formation when using only one SDI has been demonstrated up to 3 Hz and repetition rates for a full system such as used here can likely be increased significantly with improved mechanical design.
\par
A series of wedges were used to attenuate the high power laser beam after reflection off the DPM on the way to the diagnostics table. 
Two CCD cameras measured the scatter and leakage through selected wedges to determine the pointing after the DPM, the total energy throughput, and to provide a qualitative analysis of the near field beam mode. 
The camera images were calibrated using an energy meter. The beam was focused using a set of curved wedges onto another CCD camera mounted on a translation stage, providing the capability to scan through the focus. 
Discussion of the DPM reflected spatial mode refer to this measurement.
The characterization of the temporal contrast enhancement on a few 100 picosecond timescale was measured using a scanning third-order cross-correlator (Sequoia by Amplitude Technologies), which has a contrast dynamic range of at least 9 orders of magnitude and capable of measuring from ~200 ps before the pulse peak to the peak itself  with 100 fs resolution. 
The nanosecond temporal contrast enhancement was measured using two photo diodes, one with appropriate filtering to measure the peak and the other with less filtering capable of measuring any prepulse. 
The wavefront after DPM reflection was measured using a wavefront sensor (Haso by Imagine Optic). 
An important figure of merit is the total throughput of a DPM. 
The existing literature on DPMs is modest, but this is generally referred to as the ‘reflectivity’ of the DPM, in other words the ratio of the final output pulse energy to the input, and we do likewise. 
Finally, we refer to the beam with the DPM bypassed as the Reference.
\par

\section*{Results}
\subsection*{Experiment}

Figure 2 shows the DPM reflectivity during the focal scan described previously for which the intensity on both PMs varied from $10^{16}$ W/cm$^{2}$ to $10^{18}$ W/cm$^{2}$. 
An average of three shots per focal location were performed and the maximum total reflectivity was found to exceed 80\%, significantly higher than most reported DPM systems\cite{10.1063/1.2234850,Levy:07,Kim_2011,Choi_2020}. 
This is among the highest reflectivities reported even when compared to single PM systems\cite{10.1063/1.4954242,Thaury_2007,Scott_2015,Rodel_2011,Obst_2018,osti_1349103,PhysRevE.69.026402,Inoue:16,10.1063/1.1646737,osti_21272732,Nomura_2007,10.1063/1.5109683,Monot:04}. 
The dip in reflectivity when approaching peak fluences on PM 2 is a common characteristic seen in many SPM systems\cite{Obst_2018,10.1063/1.4954242,osti_1349103} and is explained later. 
The focal location ~40 mm upstream of PM 2, circled in Fig. 2a), was found to result in excellent spatial mode preservation after the DPM, as visible in Fig b) – with DPM and c) – without DPM, while still maintaining a high DPM reflectivity of >70\%.
This focal location was chosen for measurement of the DPM contrast enhancement and other laser pulse properties after DPM reflection. 
These latter measurements were performed using the curved wedges and CCD camera described earlier. 
The measurements described next were performed at this focal location. 

\par
The temporal contrast enhancement due to the DPM was measured using a scanning third order cross-correlator. 
Such measurements generally require a large number of laser shots, often taking 1-3 hours at a 1 Hz repetition rate. 
While the DPM system was not operated at 1 Hz, it is largely expected that much of the temporal contrast after the DPM would fall below the detection threshold of the Sequoia. 
As such, we limited our focus to two temporal ranges of interest: measurement of a substantial pre-pulse 167 ps before the main pulse, and the pulse pedestal starting about 20 ps before the main pulse peak (see Fig. 3). 
The measurement using the DPM was averaged over ~3 shots per time step and shots with signal below the Sequoia background level were not included in the analysis. 
We observe a contrast enhancement ranging between two and three orders of magnitude. 
The variation in contrast is likely due to the film thickness fluctuations  as well as slight variations of alignment into the very alignment sensitive Sequoia. 
The SDI wiping speeds used here typically produce LC films ranging from 20-40 nm thick. 
Estimating the reflectivity in this range due to thin film interference for each PM results in an estimated contrast enhancement factor ranging from 550 to 7800. 
Finally, a pre-pulse at 3.6 ns (outside the range of the Sequoia) was identified using photodiodes. 
After the DPM, it was indistinguishable from the background level of the measurement indicating an improvement of at least ~2 orders of magnitude. 
An 810 nm diode laser was used to measure the low field reflectivity of the DPM resulting in an estimated contrast enhancement of 350. 
Future improvements permitting data collection at the full repetition rate of the BELLA laser will provide more statistics on these measurements, but a substantial improvement in pulse contrast was achieved. 

\begin{figure}[H]
\centering
\includegraphics[width=0.75\linewidth]{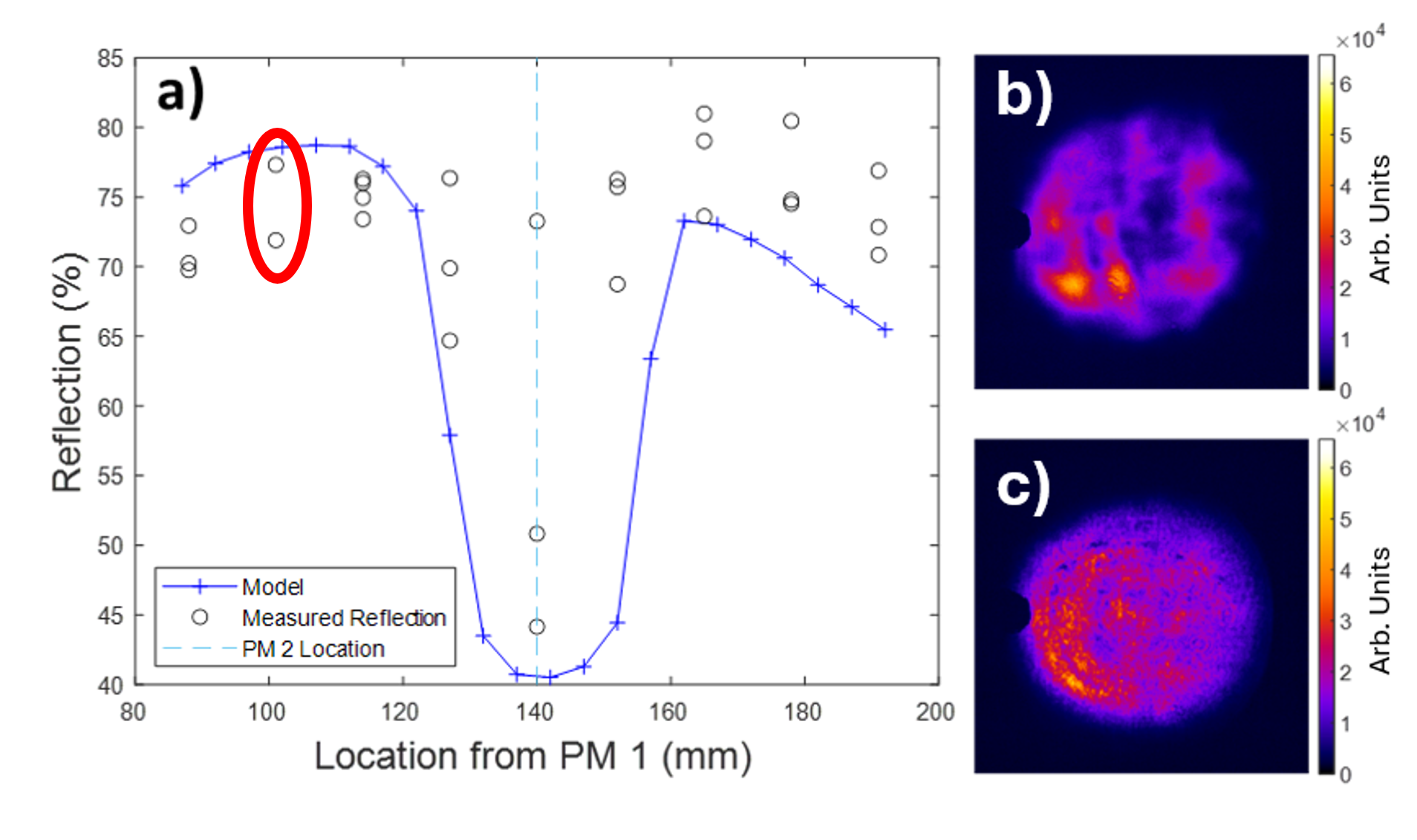}
\caption{Total reflectivity vs focal location downstream from PM 1, shown in (a), measured (black) and calculated using the model described in the text (blue). The dashed blue line indicates a focus directly on PM 2. The red circle indicates the focal location used to optimize the spatial mode and stability and for measurement of the temporal contrast enhancement. The measured Near Field reflected off the DPM (b) is similar to the Reference beam (c).}
\label{fig:Reflectivity}
\end{figure}

\begin{figure}[H]
\centering
\includegraphics[width=0.7\linewidth]{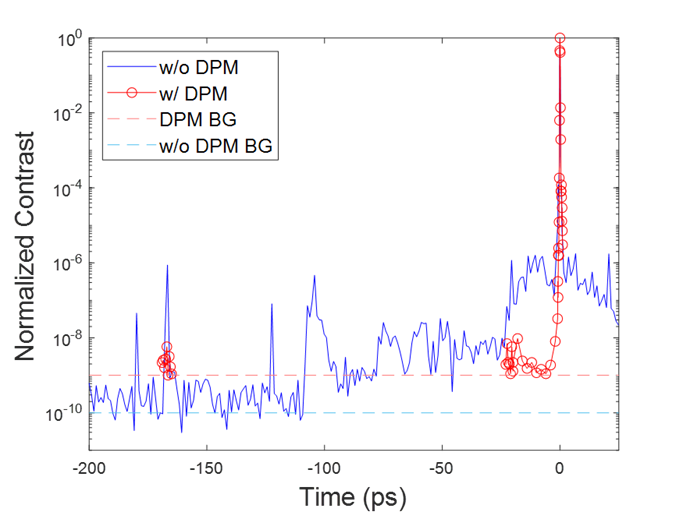}
\caption{The temporal profiles of the DPM reflected(red) and Reference (blue) pulses. The circles represent an average   taken of shots that were above the background threshold.}
\label{fig:Contrast}
\end{figure}

\par
A wavefront sensor was used to compare the Reference spatial mode quality to the DPM reflected beam. 
To our knowledge this is the first time that the wavefront of a pulse after reflection from a single- or double-PM has been reported in the literature. 
Figure 4 compares the Zernike Polynomial coefficients of the Reference beam to the DPM beam and indicates that most of the Zernike orders were largely unaffected. 
The defocus and two astigmatism coefficients were omitted from Fig. 4 as they dwarf the comparison of the higher order aberrations. 
In all three instances the coefficients were largely unchanged for the DPM beam compared to the Reference beam. 
The Root Mean Squared (RMS) and Peak-to-Valley (PV) of the DPM (Reference) wavefronts were 0.1 ± 0.02 \textmu m (0.08 ± 0.004 \textmu m) and 0.51 ± 0.079 \textmu m (0.4 ± 0.03 \textmu m), respectively. 
The RMS and PV numbers were calculated after removing the focus Zernike coefficient which showed no significant change between Reference and DPM beams. 
The Zernike coefficients stability and overall similar quality to the reference wavefront are clear indications that although the beam is reflecting from an expanding plasma surface, each PM surface remains flat for the entirety of the reflection.

\begin{figure}[H]
\centering
\includegraphics[width=0.9\linewidth]{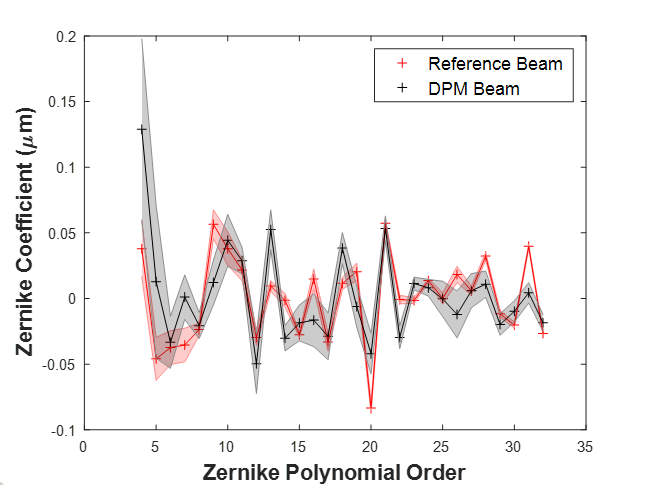}
\caption{Zernike Polynomial Coefficients of the Reference and DPM beams. The coefficients were averaged over \textasciitilde20 shots for each case. The shaded regions represent one standard deviation. The focus and astigmatism coefficients were omitted, but each showed no significant change for the DPM beam.}
\label{fig:Zernike}
\end{figure}

\par
A flat wavefront is required to create the best possible focus, however a more common metric of the focus is to simply measure the FWHM size of the beam profile. 
A comparison of the optimized DPM and Reference spatial modes is shown in Fig. 5. 
The FWHM size of the horizontal and vertical lineouts of the DPM (Reference) modes were 80 \textmu m (68 \textmu m) and 62 \textmu m (72 \textmu m), respectively, indicating good preservation of the beam mode at focus. 
The CCD camera imaging the spatial mode was translated to measure the behavior of the beam coming in and out of focus. Fig. 5 c) and d) show the FWHM size of Gaussian fits to the focus summed in the horizontal and vertical directions for the Reference and DPM beams. 
The horizontal and vertical lineouts and Gaussian fits indicate that there is no significant difference between the DPM and Reference beams, consistent with the Zernike measurements above. 

\begin{figure}[H]
\centering
\includegraphics[width=0.8\linewidth]{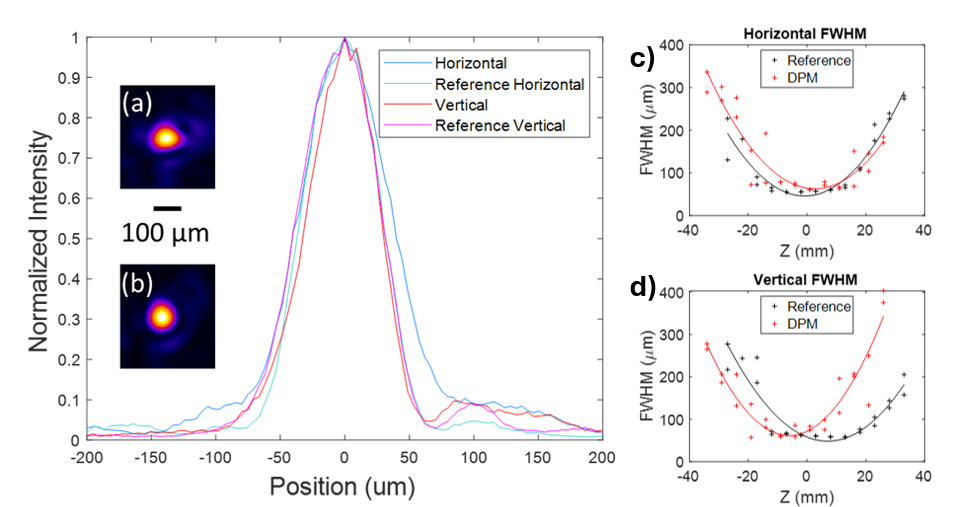}
\caption{Horizontal and vertical lineouts of an optimized spatial mode of the DPM (blue and red, respectively) and Reference (cyan and magenta, respectively) beam. The insets show the DPM (a) and Reference (b) spatial modes used for the lineouts. The FWHM of Gaussian fits of the focus summed  in the horizontal and vertical directions are shown in (c) and (d), respectively.   Parabolic fits of the FWHM are shown in (c) and (d).}
\label{fig:Focus}
\end{figure}

\par
A focal scan using a pulse energy of 20 J was also performed, but without re-optimizing the locations of PM 1 and PM 2 due to lack of time. 
Increasing the energy without changing the locations of each PM means that at no point during the 20 J scan was the laser incident on both PMs with the same incident intensity from the 7 J scans. 
The maximum total reflectivity was still measured to be >75\%, and thus still larger than that of other DPM systems reported in the literature that we are aware of. 

\subsection*{Discussion}

The physics of PM operation is still not fully understood while new uses and configurations for PMs continue to be explored\cite{Quéré_Vincenti_2021,Rodel_2011,PhysRevLett.119.094801}. 
A better understanding of PMs requires treating PM performance over several orders of magnitude in intensity from below the ionization limit to well above. 
Several groups have developed models\cite{Zingale_thesis,10.1063/1.4954242,Obst_2018,Scott_2015,10.1063/1.2234850,PhysRevE.69.026402} as well as numerical simulations\cite{10.1063/1.5109683,osti_1349103} to predict this behavior. 
However, this work has often focused on one of two regimes, either the rise in reflectivity with increased incident intensity or the sudden drop in reflectivity or dip after a material and laser dependent intensity threshold is reached. 
To compound this issue, the dominant physics at play is not fully agreed upon. 
There have been studies that indicate that the physics of PMs is dominated by incident fluence\cite{PhysRevE.69.026402,10.1063/1.2234850}, rather than incident intensity. 
While the various previous work fits their respective data well, no group has been able to accurately predict the total reflectivity regime covered by PM operation. 

\par
We present the first model to correctly describe both the reflectivity rise with intensity and the subsequent dip (see Fig. 2). 
A more detailed description is presented in the Methods section, but in broad terms, the reflectivity depends on the changing vacuum – plasma interface. The total reflectivity from a single PM system can be expressed as: 
\begin{equation}\label{PM_eqn}
    R_{\text{SPM}} = R_{\text{plas}} \times R_{\text{exp}}
\end{equation}
\(R_{\text{plas}}\) is the reflectivity from a planar plasma interface\cite{Zingale_thesis} and is the dominant factor for the rise in reflectivity of a PM as a function of intensity. 
\(R_{\text{exp}}\) accounts for the drop in specular reflectivity due to plasma mirror surface expansion from heating by the leading edge of the pulse. 
The amount of expansion is determined by the time when the laser temporal profile crosses the ionization threshold before the peak of the pulse. 
\(R_{\text{exp}}\) effectively determines the intensity at which the fall-off occurs and how sharp the fall off is\cite{Obst_2018,10.1063/1.4954242,Scott_2015,10.1063/1.325037}. 
To accurately model the DPM system used in the experiment, equation (1) is first applied to PM 1, then onto the PM 2 using the reduced pulse energy and estimated improved contrast. 
The total DPM reflectivity is the reflectivity of each PM multiplied together. 

\par
Dissecting the model’s predictions for the DPM system allows us to gain a better understanding of the plasma mirror response. 
The model estimates that the reflectivity from PM 1 reaches as high as 86\%, however the incident intensity on PM 1 is never large enough to see the characteristic reflectivity dip. 
The reflectivity from PM 2 as a function of incident intensity shows that the peak reflectivity of PM 2 alone is almost 95\% for an incident intensity of 5x10$^{17}$ W/cm$^{2}$. 
This is in stark contrast  to both the peak reflectivity and incident intensity for a single LC PM experiment previously performed at the Astra laser which had a peak intensity of ~2x10$^{16}$ W/cm$^{2}$, shown later in Fig. 6. 
As both lasers used had roughly the same pulse duration, 30 - 40 fs, this discrepancy is not explained by the pulse fluences. 
The explanation for this behavior can be found by looking at the full temporal contrast of each pulse. 
Although both pulses are below the ionization threshold for almost the entirety of their durations, the Astra laser pulse reached this threshold >1ps earlier than the BELLA laser pulse used in this work. 
This resulted in the PM surface expanding more for the Astra pulse and thus reducing the specular reflectivity from the surface, described by \(R_{\text{exp}}\). 
In both cases, the pre-pulse was determined not to play a role in the peak reflectivity, with the ionization threshold being crossed \( \lesssim \)2 ps before the peak of the pulse. 
This dependence on ionization threshold also explains why models that accurately predict the rise in reflectivity do not capture the reflectivity dip. 
In the numerical simulations, the appropriate ionization threshold may be used but the temporal pulse profile is often idealized\cite{10.1063/1.5109683,osti_1349103}. 
This works well at incident lower intensities (\textasciitilde10$^{14}$-10$^{15}$ W/cm$^{2}$), as in this regime experimental pulse shapes covering the ionization threshold to the peak intensity are more accurately represented by an idealized fit. 
However, actual temporal pulse profiles eventually deviate away from ideal shapes, hence the need for contrast measurements. 
This deviation away from an ideal shape leads to earlier surface expansions for a given peak intensities than would be seen in a simulation. 
Models based on the incident fluence would never see any effects due to ionization as this is an intensity driven process\cite{PhysRevE.69.026402}. 

\section*{Conclusion}
We have presented a highly efficient, novel DPM system utilizing free standing ultrathin LC films and implemented on the PW laser system at the BELLA Center, operated at 7 and 20 J. 
The energy throughput or total reflectivity was an impressive >80\% and is the highest seen for any reported DPM system, even exceeding that of many single PM systems. 
After optimization, it was shown that there was no significant degradation of the near field wavefront quality or focusability. 
The temporal contrast enhancement on the nano- and picosecond timescale was found to be two to three orders of magnitude. 
The ability to form films \textit{in situ} before every shot eliminates the need for large substrates that are rastered between shots, allowing for PM or DPM systems to run for extended periods of time. 
LC films have been shown to form films at up to 3 Hz using the same SDI systems used in this work, paving the way for a high repetition rate, compact DPM system for PW laser experiments. 

\par
We also present a new model that correctly predicts the increase in reflectivity with increased incident intensity, the peak reflectivity and fall-off of reflectivity for a PM or DPM system. 
To our knowledge, this is the first model to do this. 
The model was tested using two different data sets (see methods section) and performed well in both cases. 
The model predicts that the peak reflectivity and fall off is highly sensitive to the specific nature of the temporal rise of the laser pulse. 
It suggests that the level of light even \textasciitilde1 ps before the peak of the pulse plays a critical role in determining when the reflectivity will start to fall-off with intensity. 
These results will facilitate better designs of plasma mirror systems. 

\section*{Methods}
\subsection*{Plasma Mirror Reflectivity Model}
The \(R_{\text{exp}}\) term in equation \ref{PM_eqn} is described in Shaw, et al.\cite{10.1063/1.4954242}. 
It is based on a formalism suitable for treating the effect of surface roughness but, here, it is used to account for the effect of non-planar plasma expansion. 
\(R_{\text{exp}}\) is given by:
\begin{equation}\label{exp_eqn}
R_{\text{exp}} = \exp \left[ - \left( \frac{4 \pi \sigma_r}{\lambda} \cos \theta_i \right)^2 \right]
\end{equation}
where \( \sigma_r \) is a measure of the PM surface variation, and \(\theta_i\) is the angle of incidence. 
Briefly, the first step is to devise a 1+1 dimensional (1D space and time) representation of the pulse. 
For this work, a spatial Gaussian profile is combined with the measured temporal profile including pre-pulse and the measured pulse energy. 
The spatially dependent time at which the ionization threshold is crossed is then determined across the PM surface. 
The expansion of the PM is determined by multiplying the ion sound speed by the time interval between ionization and the peak of the pulse. 
The standard deviation of the variation across the PM yields \( \sigma_r \). 
For this work, the spot size on each PM surface was given by a \( \text{TEM}_{00} \) Gaussian mode with appropriate beam size depending on the location of the 60 \textmu m beam waist. 
The ion sound speed was 1.8 x 10$^{5}$ m/s \cite{Obst_2018,10.1063/1.4954242}. 
The ionization threshold was 9.4 x 10$^{13}$ W/cm$^{2}$, based on the appearance intensities for the ionization levels of 8CB\cite{10.1063/1.5109683}. 
Neither the ion sound speed nor ionization threshold were used as a fitting parameter.

\par
\(R_{\text{plas}}\) in equation \ref{PM_eqn} was established recently by us in Zingale\cite{PhysRevAccelBeams.24.121301} but a brief description is provided here. 
It is calculated with the Fresnel coefficients using a complex index of refraction based on the Drude model\cite{Born_Wolf_Bhatia_Clemmow_Gabor_Stokes_Taylor_Wayman_Wilcock_1999}, $\hat{n}$:
\begin{equation}\label{drude_eqn}
\hat{n} = n(1 + ik)
\end{equation}
where
\begin{equation}\label{ref_eqn}
n = \sqrt{\frac{1 - \alpha + \sqrt{(\alpha - 1)^2 + \alpha^2 \left(\frac{\nu_{ei}}{\omega_L}\right)^2}}{2}}
\end{equation}
and
\begin{equation}\label{imag_eqn}
k = \frac{\alpha \nu_{ei}}{2 \omega_L n^2}
\end{equation}
The parameter $\alpha$ is defined as
\begin{equation}\label{alpha_eqn}
\alpha = \frac{\omega_p^2}{\omega_L^2 + \nu_{ei}^2}
\end{equation}
and \( \nu_{ei} \) is the Spitzer collision rate\cite{10.1119/1.1969155}, and $\omega_p$ and $\omega_L$ are the plasma and laser frequencies, respectively. 
The front surface temperature used to calculate \( \nu_{ei} \) is given by Gibbon\cite{Short_Pulse_Gibbon} and its dependence is:
\begin{equation}\label{temp_eqn}
T_e \propto \left[\frac{(\eta I_0)^2 Z t}{n_e}\right]^{\frac{2}{9}}
\end{equation}
where \( \eta \) is the fractional absorption, \( I_0 \) is the incident intensity, \( Z \) is the ionization state (chosen to be 4, based on the intensity range to be explored), \( t \) is the FWHM pulse duration, and \( n_e \) is the electron density. 
The initial electron density is \( 2.4 \times 10^{23} \, \text{cm}^{-3} \) for 8CB and \( Z = 4 \). 
Given an initial guess for the absorption \( \eta \), equations \ref{drude_eqn}-\ref{temp_eqn} are evaluated to determine the Fresnel reflectance and transmittance and, thus, the absorption. 
The calculation is then repeated using the newly obtained absorption for \( \eta \) until convergence in the value of \( \eta \) is achieved, usually requiring only several iterations.

\par
The model is designed to estimate the reflection from a single LC PM. 
To estimate the total reflectivity using two PMs, the estimated reflectivity of each PM are multiplied together. 
Additionally, the model uses an energy incident on PM 2 calculated from multiplying the original incident energy on PM 1 multiplied by the estimated reflectivity of PM 1. 
The temporal contrast incident on PM 2 is estimated to be improved from the pulse cleaning effects of PM 1. 
In this way the parameters inherent to the LC are unchanged but the change due to incident experimental parameters have been accounted for. 

\par
This model was tested using results from a previous experiment characterizing a single LC PM performed on the Astra laser at the Central Laser Facility\cite{osti_1349103} (see Fig. 6). 
The only changes  made to model were inherent to the laser used, such as incident energy and replacing the measured temporal profile of the BELLA laser with the temporal profile of the Astra laser\cite{Tang:14}. 
Also shown is a previously reported\cite{10.1063/1.5109683} particle-in-cell (PIC) simulation modeling the PM evolution from a cold, neutral substrate to a plasma including varying permittivity and ionization state that correctly described the reflectivity up to an intensity of \( 2 \times 10^{16} \, \text{W/cm}^2 \), however it failed to accurately predict the reflectivity peak and resulting dip in reflectivity. 
Crucially, the simulation did not include the pulse pedestal and thus the resulting PM expansion. 
The new model described above accurately predicts the reflectivity over two orders of magnitude of incident intensity and, importantly, correctly describes both the regimes of the rise of reflectivity with increased intensity and the overall peak reflectivity with the reflectivity drop off. 
To our knowledge, this is the first model that can reliably make predictions in both regimes. 

\begin{figure}[H]
\centering
\includegraphics[width=0.7\linewidth]{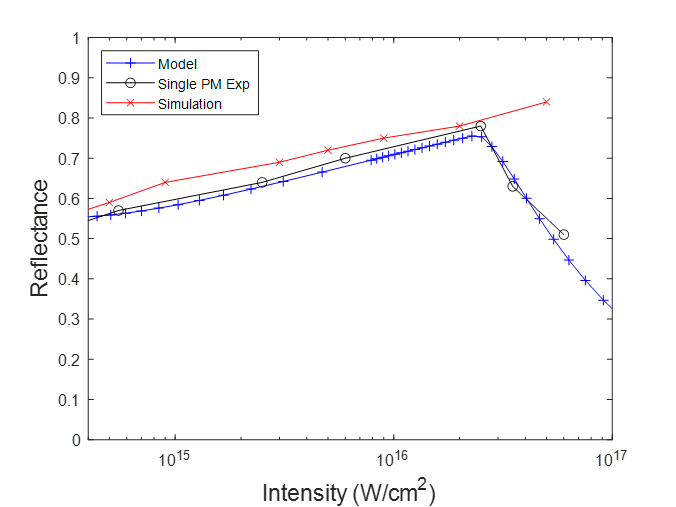}
\caption{Reflectivity vs Intensity as calculated by the model described in the text (blue) compared to the measured reflection (black) and a PIC simulation (red).}
\label{fig:Validation}
\end{figure}

\bibliography{DPM_Ref}

\begin{thebibliography}{10}
\urlstyle{rm}
\expandafter\ifx\csname url\endcsname\relax
  \def\url#1{\texttt{#1}}\fi
\expandafter\ifx\csname urlprefix\endcsname\relax\def\urlprefix{URL }\fi
\expandafter\ifx\csname doiprefix\endcsname\relax\def\doiprefix{DOI: }\fi
\providecommand{\bibinfo}[2]{#2}
\providecommand{\eprint}[2][]{\url{#2}}

\bibitem{Yoon:21}
\bibinfo{author}{Yoon, J.~W.} \emph{et~al.}
\newblock \bibinfo{journal}{\bibinfo{title}{Realization of laser intensity over $>10^{23}$ w/cm$^{2}$}}.
\newblock {\emph{\JournalTitle{Optica}}} \textbf{\bibinfo{volume}{8}}, \bibinfo{pages}{630--635}, \doiprefix\url{10.1364/OPTICA.420520} (\bibinfo{year}{2021}).

\bibitem{Hadjisolomou:10.1063/1.5124457}
\bibinfo{author}{Hadjisolomou, P.} \emph{et~al.}
\newblock \bibinfo{journal}{\bibinfo{title}{{Preplasma effects on laser ion generation from thin foil targets}}}.
\newblock {\emph{\JournalTitle{Physics of Plasmas}}} \textbf{\bibinfo{volume}{27}}, \bibinfo{pages}{013107}, \doiprefix\url{10.1063/1.5124457} (\bibinfo{year}{2020}).

\bibitem{Hegelich_2013}
\bibinfo{author}{Hegelich, B.~M.} \emph{et~al.}
\newblock \bibinfo{journal}{\bibinfo{title}{Laser-driven ion acceleration from relativistically transparent nanotargets}}.
\newblock {\emph{\JournalTitle{New Journal of Physics}}} \textbf{\bibinfo{volume}{15}}, \bibinfo{pages}{085015}, \doiprefix\url{10.1088/1367-2630/15/8/085015} (\bibinfo{year}{2013}).

\bibitem{10.1063/1.2220011}
\bibinfo{author}{Neely, D.} \emph{et~al.}
\newblock \bibinfo{journal}{\bibinfo{title}{{Enhanced proton beams from ultrathin targets driven by high contrast laser pulses}}}.
\newblock {\emph{\JournalTitle{Applied Physics Letters}}} \textbf{\bibinfo{volume}{89}}, \bibinfo{pages}{021502}, \doiprefix\url{10.1063/1.2220011} (\bibinfo{year}{2006}).

\bibitem{10.1063/1.2480610}
\bibinfo{author}{Antici, P.} \emph{et~al.}
\newblock \bibinfo{journal}{\bibinfo{title}{{Energetic protons generated by ultrahigh contrast laser pulses interacting with ultrathin targets}}}.
\newblock {\emph{\JournalTitle{Physics of Plasmas}}} \textbf{\bibinfo{volume}{14}}, \bibinfo{pages}{030701}, \doiprefix\url{10.1063/1.2480610} (\bibinfo{year}{2007}).

\bibitem{PhysRevLett.99.185002}
\bibinfo{author}{Ceccotti, T.} \emph{et~al.}
\newblock \bibinfo{journal}{\bibinfo{title}{Proton acceleration with high-intensity ultrahigh-contrast laser pulses}}.
\newblock {\emph{\JournalTitle{Phys. Rev. Lett.}}} \textbf{\bibinfo{volume}{99}}, \bibinfo{pages}{185002}, \doiprefix\url{10.1103/PhysRevLett.99.185002} (\bibinfo{year}{2007}).

\bibitem{RevModPhys.85.751}
\bibinfo{author}{Macchi, A.}, \bibinfo{author}{Borghesi, M.} \& \bibinfo{author}{Passoni, M.}
\newblock \bibinfo{journal}{\bibinfo{title}{Ion acceleration by superintense laser-plasma interaction}}.
\newblock {\emph{\JournalTitle{Rev. Mod. Phys.}}} \textbf{\bibinfo{volume}{85}}, \bibinfo{pages}{751--793}, \doiprefix\url{10.1103/RevModPhys.85.751} (\bibinfo{year}{2013}).

\bibitem{Zingale_thesis}
\bibinfo{author}{Zingale, A.}
\newblock \bibinfo{journal}{\bibinfo{title}{Optical response of plasmas from moderate intensity to the relativistic regime}}.
\newblock {\emph{\JournalTitle{Ohio State University, Doctoral dissertation}}}  (\bibinfo{year}{2021}).

\bibitem{Pozderac_thesis}
\bibinfo{author}{Pozderac, P.}
\newblock \bibinfo{journal}{\bibinfo{title}{Novel pump-probe particle-in-cell simulations of relativistic transparency and birefringence}}.
\newblock {\emph{\JournalTitle{Ohio State University, Doctoral dissertation}}}  (\bibinfo{year}{2022}).

\bibitem{10.1063/1.4954242}
\bibinfo{author}{Shaw, B.~H.}, \bibinfo{author}{Steinke, S.}, \bibinfo{author}{van Tilborg, J.} \& \bibinfo{author}{Leemans, W.~P.}
\newblock \bibinfo{journal}{\bibinfo{title}{{Reflectance characterization of tape-based plasma mirrors}}}.
\newblock {\emph{\JournalTitle{Physics of Plasmas}}} \textbf{\bibinfo{volume}{23}}, \bibinfo{pages}{063118}, \doiprefix\url{10.1063/1.4954242} (\bibinfo{year}{2016}).

\bibitem{Thaury_2007}
\bibinfo{author}{Thaury, C.} \emph{et~al.}
\newblock \bibinfo{journal}{\bibinfo{title}{Plasma mirrors for ultrahigh-intensity optics}}.
\newblock {\emph{\JournalTitle{Nature Physics}}} \textbf{\bibinfo{volume}{3}}, \bibinfo{pages}{424--429}, \doiprefix\url{10.1038/Nphys595} (\bibinfo{year}{2007}).

\bibitem{Scott_2015}
\bibinfo{author}{Scott, G.~G.} \emph{et~al.}
\newblock \bibinfo{journal}{\bibinfo{title}{Optimization of plasma mirror reflectivity and optical quality using double laser pulses}}.
\newblock {\emph{\JournalTitle{New Journal of Physics}}} \textbf{\bibinfo{volume}{17}}, \bibinfo{pages}{033027}, \doiprefix\url{10.1088/1367-2630/17/3/033027} (\bibinfo{year}{2015}).

\bibitem{Rodel_2011}
\bibinfo{author}{Rödel, C.} \emph{et~al.}
\newblock \bibinfo{journal}{\bibinfo{title}{High repetition rate plasma mirror for temporal contrast enhancement of terawatt femtosecond laser pulses by three orders of magnitude}}.
\newblock {\emph{\JournalTitle{Applied Physics B}}} \textbf{\bibinfo{volume}{103}}, \bibinfo{pages}{295--302}, \doiprefix\url{10.1007/s00340-010-4329-7} (\bibinfo{year}{2011}).

\bibitem{Obst_2018}
\bibinfo{author}{Obst, L.} \emph{et~al.}
\newblock \bibinfo{journal}{\bibinfo{title}{On-shot characterization of single plasma mirror temporal contrast improvement}}.
\newblock {\emph{\JournalTitle{Plasma Physics and Controlled Fusion}}} \textbf{\bibinfo{volume}{60}}, \bibinfo{pages}{054007}, \doiprefix\url{10.1088/1361-6587/aab3bb} (\bibinfo{year}{2018}).

\bibitem{osti_1349103}
\bibinfo{author}{Poole, P.~L.} \emph{et~al.}
\newblock \bibinfo{journal}{\bibinfo{title}{Experiment and simulation of novel liquid crystal plasma mirrors for high contrast, intense laser pulses}}.
\newblock {\emph{\JournalTitle{Scientific Reports}}} \textbf{\bibinfo{volume}{6}}, \doiprefix\url{10.1038/srep32041} (\bibinfo{year}{2016}).

\bibitem{PhysRevE.69.026402}
\bibinfo{author}{Doumy, G.} \emph{et~al.}
\newblock \bibinfo{journal}{\bibinfo{title}{Complete characterization of a plasma mirror for the production of high-contrast ultraintense laser pulses}}.
\newblock {\emph{\JournalTitle{Phys. Rev. E}}} \textbf{\bibinfo{volume}{69}}, \bibinfo{pages}{026402}, \doiprefix\url{10.1103/PhysRevE.69.026402} (\bibinfo{year}{2004}).

\bibitem{Inoue:16}
\bibinfo{author}{Inoue, S.} \emph{et~al.}
\newblock \bibinfo{journal}{\bibinfo{title}{Single plasma mirror providing 104 contrast enhancement and 70\% reflectivity for intense femtosecond lasers}}.
\newblock {\emph{\JournalTitle{Appl. Opt.}}} \textbf{\bibinfo{volume}{55}}, \bibinfo{pages}{5647--5651}, \doiprefix\url{10.1364/AO.55.005647} (\bibinfo{year}{2016}).

\bibitem{10.1063/1.1646737}
\bibinfo{author}{Dromey, B.}, \bibinfo{author}{Kar, S.}, \bibinfo{author}{Zepf, M.} \& \bibinfo{author}{Foster, P.}
\newblock \bibinfo{journal}{\bibinfo{title}{The plasma mirror—a subpicosecond optical switch for ultrahigh power lasers}}.
\newblock {\emph{\JournalTitle{Review of Scientific Instruments}}} \textbf{\bibinfo{volume}{75}}, \bibinfo{pages}{645--649}, \doiprefix\url{10.1063/1.1646737} (\bibinfo{year}{2004}).

\bibitem{osti_21272732}
\bibinfo{author}{Yi, C.} \emph{et~al.}
\newblock \bibinfo{journal}{\bibinfo{title}{Time-resolved measurements on reflectivity of an ultrafast laser-induced plasma mirror}}.
\newblock {\emph{\JournalTitle{Physics of Plasmas}}} \textbf{\bibinfo{volume}{16}}, \doiprefix\url{10.1063/1.3247865} (\bibinfo{year}{2009}).

\bibitem{Nomura_2007}
\bibinfo{author}{Nomura, Y.} \emph{et~al.}
\newblock \bibinfo{journal}{\bibinfo{title}{Time-resolved reflectivity measurements on a plasma mirror with few-cycle laser pulses}}.
\newblock {\emph{\JournalTitle{New Journal of Physics}}} \textbf{\bibinfo{volume}{9}}, \bibinfo{pages}{9}, \doiprefix\url{10.1088/1367-2630/9/1/009} (\bibinfo{year}{2007}).

\bibitem{10.1063/1.5109683}
\bibinfo{author}{Cochran, G.~E.}, \bibinfo{author}{Poole, P.~L.} \& \bibinfo{author}{Schumacher, D.~W.}
\newblock \bibinfo{journal}{\bibinfo{title}{Modeling pulse-cleaning plasma mirrors from dielectric response to saturation: A particle-in-cell approach}}.
\newblock {\emph{\JournalTitle{Physics of Plasmas}}} \textbf{\bibinfo{volume}{26}}, \bibinfo{pages}{103103}, \doiprefix\url{10.1063/1.5109683} (\bibinfo{year}{2019}).

\bibitem{Monot:04}
\bibinfo{author}{Monot, P.} \emph{et~al.}
\newblock \bibinfo{journal}{\bibinfo{title}{High-order harmonic generation by nonlinear reflection of an intense high-contrast laser pulse on a plasma}}.
\newblock {\emph{\JournalTitle{Opt. Lett.}}} \textbf{\bibinfo{volume}{29}}, \bibinfo{pages}{893--895}, \doiprefix\url{10.1364/OL.29.000893} (\bibinfo{year}{2004}).

\bibitem{10.1063/1.2234850}
\bibinfo{author}{Wittmann, T.} \emph{et~al.}
\newblock \bibinfo{journal}{\bibinfo{title}{{Towards ultrahigh-contrast ultraintense laser pulses—complete characterization of a double plasma-mirror pulse cleaner}}}.
\newblock {\emph{\JournalTitle{Review of Scientific Instruments}}} \textbf{\bibinfo{volume}{77}}, \bibinfo{pages}{083109}, \doiprefix\url{10.1063/1.2234850} (\bibinfo{year}{2006}).

\bibitem{Levy:07}
\bibinfo{author}{L\'{e}vy, A.} \emph{et~al.}
\newblock \bibinfo{journal}{\bibinfo{title}{Double plasma mirror for ultrahigh temporal contrast ultraintense laser pulses}}.
\newblock {\emph{\JournalTitle{Opt. Lett.}}} \textbf{\bibinfo{volume}{32}}, \bibinfo{pages}{310--312}, \doiprefix\url{10.1364/OL.32.000310} (\bibinfo{year}{2007}).

\bibitem{Kim_2011}
\bibinfo{author}{Kim, I.~J.} \emph{et~al.}
\newblock \bibinfo{journal}{\bibinfo{title}{Spatio-temporal characterization of double plasma mirror for ultrahigh contrast and stable laser pulse}}.
\newblock {\emph{\JournalTitle{Applied Physics B}}} \textbf{\bibinfo{volume}{104}}, \bibinfo{pages}{81--86}, \doiprefix\url{10.1007/s00340-011-4584-2} (\bibinfo{year}{2011}).

\bibitem{PhysRevAccelBeams.24.121301}
\bibinfo{author}{Zingale, A.} \emph{et~al.}
\newblock \bibinfo{journal}{\bibinfo{title}{Emittance preserving thin film plasma mirrors for gev scale laser plasma accelerators}}.
\newblock {\emph{\JournalTitle{Phys. Rev. Accel. Beams}}} \textbf{\bibinfo{volume}{24}}, \bibinfo{pages}{121301}, \doiprefix\url{10.1103/PhysRevAccelBeams.24.121301} (\bibinfo{year}{2021}).

\bibitem{7934119}
\bibinfo{author}{Nakamura, K.} \emph{et~al.}
\newblock \bibinfo{journal}{\bibinfo{title}{Diagnostics, control and performance parameters for the bella high repetition rate petawatt class laser}}.
\newblock {\emph{\JournalTitle{IEEE Journal of Quantum Electronics}}} \textbf{\bibinfo{volume}{53}}, \bibinfo{pages}{1--21}, \doiprefix\url{10.1109/JQE.2017.2708601} (\bibinfo{year}{2017}).

\bibitem{Schumacher_2017}
\bibinfo{author}{Schumacher, D.} \emph{et~al.}
\newblock \bibinfo{journal}{\bibinfo{title}{Liquid crystal targets and plasma mirrors for laser based ion acceleration}}.
\newblock {\emph{\JournalTitle{Journal of Instrumentation}}} \textbf{\bibinfo{volume}{12}}, \bibinfo{pages}{C04023}, \doiprefix\url{10.1088/1748-0221/12/04/C04023} (\bibinfo{year}{2017}).

\bibitem{Choi_2020}
\bibinfo{author}{Choi, I.~W.} \emph{et~al.}
\newblock \bibinfo{journal}{\bibinfo{title}{Highly efficient double plasma mirror producing ultrahigh-contrast multi-petawatt laser pulses}}.
\newblock {\emph{\JournalTitle{Opt. Lett.}}} \textbf{\bibinfo{volume}{45}}, \bibinfo{pages}{6342--6345}, \doiprefix\url{10.1364/OL.409749} (\bibinfo{year}{2020}).

\bibitem{Quéré_Vincenti_2021}
\bibinfo{author}{Quéré, F.} \& \bibinfo{author}{Vincenti, H.}
\newblock \bibinfo{journal}{\bibinfo{title}{Reflecting petawatt lasers off relativistic plasma mirrors: a realistic path to the schwinger limit}}.
\newblock {\emph{\JournalTitle{High Power Laser Science and Engineering}}} \textbf{\bibinfo{volume}{9}}, \bibinfo{pages}{e6}, \doiprefix\url{10.1017/hpl.2020.46} (\bibinfo{year}{2021}).

\bibitem{PhysRevLett.119.094801}
\bibinfo{author}{Za\"{\i}m, N.}, \bibinfo{author}{Th\'evenet, M.}, \bibinfo{author}{Lifschitz, A.} \& \bibinfo{author}{Faure, J.}
\newblock \bibinfo{journal}{\bibinfo{title}{Relativistic acceleration of electrons injected by a plasma mirror into a radially polarized laser beam}}.
\newblock {\emph{\JournalTitle{Phys. Rev. Lett.}}} \textbf{\bibinfo{volume}{119}}, \bibinfo{pages}{094801}, \doiprefix\url{10.1103/PhysRevLett.119.094801} (\bibinfo{year}{2017}).

\bibitem{10.1063/1.325037}
\bibinfo{author}{Holzer, J.~A.} \& \bibinfo{author}{Sung, C.~C.}
\newblock \bibinfo{journal}{\bibinfo{title}{Scattering of electromagnetic waves from a rough surface. ii}}.
\newblock {\emph{\JournalTitle{Journal of Applied Physics}}} \textbf{\bibinfo{volume}{49}}, \bibinfo{pages}{1002--1011}, \doiprefix\url{10.1063/1.325037} (\bibinfo{year}{1978}).

\bibitem{Born_Wolf_Bhatia_Clemmow_Gabor_Stokes_Taylor_Wayman_Wilcock_1999}
\bibinfo{author}{Born, M.} \emph{et~al.}
\newblock \emph{\bibinfo{title}{Principles of Optics: Electromagnetic Theory of Propagation, Interference and Diffraction of Light}} (\bibinfo{publisher}{Cambridge University Press}, \bibinfo{year}{1999}), \bibinfo{edition}{7} edn.

\bibitem{10.1119/1.1969155}
\bibinfo{author}{Spitzer, L.} \& \bibinfo{author}{Seeger, R.~J.}
\newblock \bibinfo{journal}{\bibinfo{title}{Physics of fully ionized gases}}.
\newblock {\emph{\JournalTitle{American Journal of Physics}}} \textbf{\bibinfo{volume}{31}}, \bibinfo{pages}{890--891}, \doiprefix\url{10.1119/1.1969155} (\bibinfo{year}{1963}).

\bibitem{Short_Pulse_Gibbon}
\bibinfo{author}{Gibbon, P.}
\newblock \emph{\bibinfo{title}{Short Pulse Laser Interactions with Matter: An Introduction}} (\bibinfo{publisher}{Imperial College Press}, \bibinfo{year}{2005}).

\bibitem{Tang:14}
\bibinfo{author}{Tang, Y.} \emph{et~al.}
\newblock \bibinfo{journal}{\bibinfo{title}{Transmission grating stretcher for contrast enhancement of high power lasers}}.
\newblock {\emph{\JournalTitle{Opt. Express}}} \textbf{\bibinfo{volume}{22}}, \bibinfo{pages}{29363--29374}, \doiprefix\url{10.1364/OE.22.029363} (\bibinfo{year}{2014}).

\end{thebibliography}

\section*{Acknowledgements}

The authors thank the BELLA PW laser team for their assistance. This work was supported by the U.S. Department of Energy Office of Science Offices of High Energy Physics, Fusion Energy Sciences, and LaserNetUS (lasernetus.org) under Contract No. DE-AC02-05CH11231.

\section*{Author contributions statement}

N.C, D.M.N, A.Z., D.W.S. conceived of the experiment. N.C. and D.M.N. designed and implemented the experimental setup at the BELLA PW Laser with assistance from L.O-H., J.B., A.J.G., S.S., K.N. N.C. analyzed the experimental data and wrote the manuscript. C.B.S., E.E., C.G.R.G. provided the beam time and experimental and analytical discussion. All authors reviewed the manuscript.

\section*{Additional information}

\textbf{Competing financial interests:} The authors declare no competing financial interests.

\end{document}